\documentclass[a4paper,fleqn]{cas-sc}

\usepackage{bigdelim}
\usepackage[authoryear,longnamesfirst]{natbib}
\usepackage{float}
\usepackage{adjustbox}

\def\tsc#1{\csdef{#1}{\textsc{\lowercase{#1}}\xspace}}
\tsc{WGM}
\tsc{QE}
\tsc{EP}
\tsc{PMS}
\tsc{BEC}
\tsc{DE}

\begin{document}
\let\WriteBookmarks\relax
\def\floatpagepagefraction{1}
\def\textpagefraction{.001}
\shorttitle{Audio Spoof Detection with GaborNet}
\shortauthors{Waldemar Maciejko}

\title [mode = title]{Audio Spoof Detection with GaborNet}                      



\author{Waldemar Maciejko}[type=editor,
                        auid=000,
                        bioid=1,
                        prefix=PhD,
                        orcid=0009-0008-2355-3042]


\begin{abstract}
An direction of development in the extraction of features from audio signals is based on processing raw samples in the time domain. Such an approach appears to be effective, especially in the era of neural networks. An example is \verb+SincNet+. In this solution, the core of the neural network layer is a set of \verb+sinc functions+ that are convolved with the input signal. Due to the finite length of \verb+sinc functions+, distortions appear in the frequency domain of the convolved signal, the same as in the case of windowing the signal. Recently, a new approach has been developed that uses \verb+Gabor filters+ to replace \verb+sinc functions+. Due to the complex results, further modifications had to be applied, such as squared modulus or Gaussian Lowpass Pooling. In this work, an ingestion layer based on a bank of \verb+Gabor filters+, named \verb+GaborNet+, and its modifications are intensively examined within the popular RawNet2 and RawGAT-ST architectures. These have been developed for the purpose of \verb+audio spoof detection+. Another issue that has been investigated was \verb+audio augmentation+ using codec conversions, room responses, and additive noises.
\end{abstract}



\maketitle

\section{Introduction}

Modern speech synthesis (TTS) or voice conversion (VC) algorithms effectively learn the target speaker's spectral characteristics. While these algorithms have various positive uses, they can also be used to spoof audio or spread disinformation. This presents a potential vulnerability for both media users and automated speaker recognition systems as indicated by Gupta \citep{Gupta2024Vulnerability}. This reiterates the need for the continued development of algorithms to detect audio spoofing.

Since the advent of neural networks, the methods of extracting information from audio signals have been significantly transformed. Throughout the progression of audio classification, approaches that employ different variations of time-frequency features, derived from digital signal processing transformations, have emerged as a traditional means of audio representation. The most widely used features based on mel-spectrograms are computed using the squared modulus of the short-term Fourier transform (STFT). Subsequently, the spectrum gets filtered through a bank of triangular bandpass filters arranged on the mel scale (which is almost logarithmic). This process imitates the non-linear way humans perceive sound frequency \cite{1163420}.

This kind of features seems insufficient in the age of neural networks. Specifically, as input features for Convolutional Neural Networks (CNNs), they face several problems. Fixed spectral regions limit the input layer's view due to the Fourier Transform rule's defined frequency bands (index of band × sampling rate/window length). Furthermore, this method smooths the speech spectrum, which may prevent the extraction of essential narrow-band features, resulting in an inadequate signal representation \cite{8639585}.

A novel tactic that works directly on the waveform has recently been introduced. In this advanced methodology, the parameters of the filter bank can be learned, similar to the weights of a neural network layer, which are updated with each iteration of the training loop based on the computed gradient of the loss function. As a result, the feature extraction stage has evolved to become the initial layer of the neural network. Among the best-documented techniques that use learnable parameters is SincNet \cite{8639585}.  This architecture has been tested in many tasks, such as speaker recognition \cite{8639585}\cite{Jung2019RawNetAE} and audio spoof detection \cite{9414234}\cite{Tak2021EndtoEndSG}\cite{Tak2021GraphAN}. 
Another more sophisticated solution named the Learnable Frontend for Audio Classification (LEAF) has shown efficacy in multiple fields like language, speaker, and emotion recognition. However, research in its application to audio spoof detection is notably lacking. In this study, LEAF has been thoroughly assessed for audio spoof detection, presenting a suitable alternative to SincNet. The Gabor filter bank, a crucial component of LEAF that can be learned, is referred to as GaborNet in comparison to SincNet.
This paper incorporates various versions of ingestion layers into the newly developed RawNet2 and RawGAT-ST deep neural networks. It provides an evaluation of these networks under complex conditions.
Another factor that was scrutinized was data augmentation, a prevalent technique employed during the training phase that contributes to the development of a generalized model. An appropriate Machine Learning algorithm is able to learn the internal representation of objects when the training dataset is sufficiently representative. In the case of audio, the key factors include bandwidth, codec distortions, and acoustic environment. A neural network cannot extrapolate such conditions to test samples if they are significantly different from training examples. Classification systems based on Deep Neural Networks, working on real live audio data, face the following challenges: lossy compression (selected spectrum with limited range and nonlinear quantization) and channel effects (packet loss, noise).

The remainder of this paper is structured as follows. Section 2 discusses related works. Section 3 details the two architectures explored in this paper, RawNet2 and RawGAT-ST, and explains their modifications with different versions of LEAF serving as the ingestion layer. Section 4 presents detailed information about the experiments conducted, their results, and the conclusions drawn from them.

\section{Related works}

The RawNet architecture was developed to enhance data extraction from audio. It utilized a conventional convolutional layer with 128 filters directly applied to raw audio. This kind of approach was used for text-independent speaker recognition. In this solution, all parameters of filters are learned during the training stage \cite{Jung2019RawNetAE}. Higher layers utilize Residual Blocks (ResNet), which use skip connections to stabilize the training process and extract more discriminative knowledge. Another improvement was the use of identity mapping \cite{He2016IdentityMI}, as well as Gated Recurrent Units (GRU) to aggregate frame-level features into a single utterance-level embedding \cite{Jung2019RawNetAE}. These strategies indicate that directly learning discriminative knowledge from raw audio using convolutional neural network layers (CNN) can be more effective for speaker recognition than using hand-engineered spectral features. Mirco Ravanelli and Yoshua Bengio introduced a new kind of one-dimensional convolutional layer named SincNet, where the first layer performs a set of time-domain convolutions between the framed input audio and a predefined kernel function \cite{8639585}. Convolution in the time domain corresponds to filtration in the frequency domain. The Fourier Transform of a normalized sinc function has a rectangular characteristic. Deploying a band-pass filter bank with rectangular characteristics leads to separation of the input signal into multiple components. Each filter in the bank is defined by learnable $f_{1}$ (low) and $f_{2}$ (high) cut-off frequencies. Learnable parameters can be initialized based on values allocated on the mel-scale. It has been proven that SincNet leads to faster convergence of learning curves and limits the number of parameters to learn in the first convolutional layer \cite{8639585}.

Another phase in the evolution of the RawNet architecture involved applying higher blocks of ResNets and a technique called Feature Map Scaling (FMS). FMS, a form of Attention Mechanism, aims to choose only those filters that convey significant and distinguishing information \cite{Jung2020ImprovedRW}.

The use of SincNet as an ingestion layer in an application developed for audio spoofing detection was first presented by H. Tak et al. \cite{9414234}. The authors noticed that, at that time, systems had poor capability to detect the A17 type of attack (the Voice Conversion system was judged to have the highest spoofing capability in the Voice Conversion Challenge 2018) \cite{WANG2020101114}. A common countermeasure for difficult-to-detect spoofed audio is to train a classifier with data from an evaluation set that contains samples of such attacks \cite{9414234}. Nevertheless, the mechanism of retraining the classifier with newly discovered spoofing methods is not effective. The direction of development for spoofing detection methods should endeavor toward generalization to protect against unseen attacks. This assumption was the foundation for applying SincNet as a solution to increase the reliability of spoofing detection by using a high-resolution spectral front-end. However, the authors proposed sevreal modifications: fixing the raw waveform input duration to 4 seconds (equivalent to ~64,000 samples at a 16,000 Hz sampling rate), restricting the SincNet layer's filter length to 129, utilizing a larger 512 kernel for the second residual block, and applying a modified FMS. The last significant modification to the RawNet architecture was the incorporation of a Gated Recurrent Unit (GRU) layer. This was used to aggregate frame-level representations into a single utterance-level representation \cite{Jung2020ImprovedRW}. The final layer involves fully connected layers with a Softmax function, producing a two-state result: spoof versus bonafide \cite{9414234}.

It is worth mentioning a solution that treats the SincNet layer as a special case of raw audio analysis, which uses the cardinal sinc function \cite{Loweimi2019OnLI}. The authors proposed using three alternatives to the sinc function: squared-sinc, gammatone, and Gaussian kernels. They also suggested considering only the real part of the convolution outcome. Compared to the classical SincNet, the applied Gaussian function led to slightly lower errors \cite{Loweimi2019OnLI}.

The following important step in advancing anti-spoofing technology was an approach akin to Graph Attention Networks (GAT)\cite{Tak2021EndtoEndSG}\cite{Tak2021GraphAN}. This strategy operates on raw data using the SincNet layer as well. It aims to shape relationship between the spectral (GAT-S) and temporal (GAT-T) domains of the signal by aggregating discriminative information extracted from these two complementary domains \cite{Tak2021GraphAN}. Both temporal and spectral modelling is conducted separately and the resulting pair of outcomes are subsequently utilized as the input for the following fusion layer. These three parts contain GATs and Graph Pooling blocks. The findings suggest that this strategy enhances the differentiation between genuine and manipulated speech signals \cite{Tak2021EndtoEndSG}.

Ravanelli's work has been further extended by applying Gabor Filters as an alternative to the sinc function. The Fourier Transform of the sinc function is ideally effective only when its length is infinite \cite{Alan2010Prentice}. However, practical application necessitates limiting the duration of this function. Applying a finite length to the sinc function is tantamount to windowing, which can result in high frequency leakage and increased ripples in the pass-band and stop-band \cite{Alan2010Prentice}. One of the advantages of incorporating Gabor filters is the elimination of the requirement for using a window function. In SincNet, there are two trainable parameters for each filter. These are the cut-off frequencies, and their adjustability significantly reduces the number of parameters that need to be learned compared to a conventional first-layer CNN.
Noé et al. proposed replacing filters based on the sinc function with a complex-valued Gabor (CV) function to improve time-frequency resolution. Their model began with a Complex-Valued Convolutional Neural Network (CVCNN), followed by a Complex-Valued Multilayer Perceptron (CMLP). Across all layers, they employed the ReLU activation function. Experimental results demonstrated that the CVCNN's performance was on par with other top-performing systems that work on raw waveforms \cite{9054220}.

Applying signal windowing in SincNet can deform learned features, particularly around abrupt signal changes. To mitigate this effect, SincNet typically uses a short analysis window of about 25 ms. However, such short windows make it difficult to capture long-term structure and may degrade performance on classification tasks \cite{6822556}. To counteract these drawbacks, another approach named LEAF has been proposed. LEAF employs a parameterized complex-valued input filtering layer based on fully learnable Gabor filters \cite{Zeghidour2021LEAFAL}. In this method, a mel filter bank is used to initialize complex, analytic Gabor wavelets. Experimental results indicate that LEAF can outperform other approaches on tasks such as speech command recognition and language identification, while achieving comparable performance with fewer model parameters.

Because ensuring sufficient variety in a training set is not always feasible, simulating diverse conditions can be a useful alternative. One approach is RawBoost, which operates directly on the raw waveform. It applies signal-processing techniques to introduce distortions such as harmonics, multi-band filtering, signal-dependent impulses, and white noise \cite{9746213}. RawBoost does not require additional data sources, such as noise recordings or room impulse responses. Another approach relies on external data to simulate different acoustic conditions—such as reverberation, speaker position within a room, or added background noises—by introducing distortions through convolution \cite{Cohen2021ASO}\cite{Snyder2015MUSANAM}\cite{7953152}.

\section{Neural network architecture}

This section describes modifications to two state-of-the-art audio spoofing detection architectures, RawNet2 and RawGAT-ST \cite{9414234}\cite{Tak2021EndtoEndSG}. 

\begin{figure}[h]
    \centering
        \includegraphics[width=0.5\textwidth]{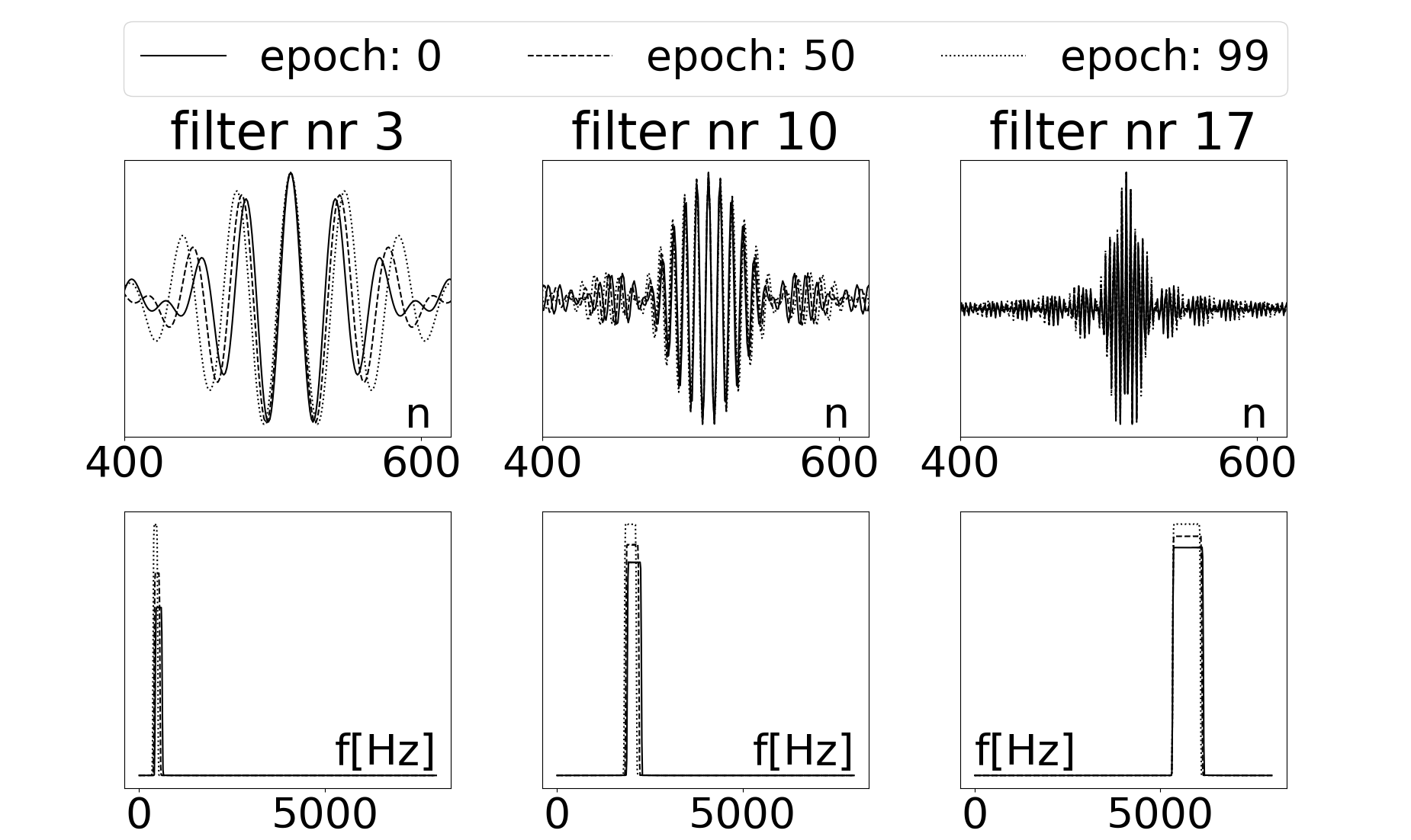}
        \caption{Three examples of sinc functions at the training stage across three epochs: $0^{th}$, $50^{th}$ and $100^{th}$. The top row presents the characteristics in the time domain, and the bottom row presents them in the frequency domain.}
        \label{fig:fig1}
\end{figure}

\begin{figure}[h]   
    \centering
        \includegraphics[width=0.5\textwidth]{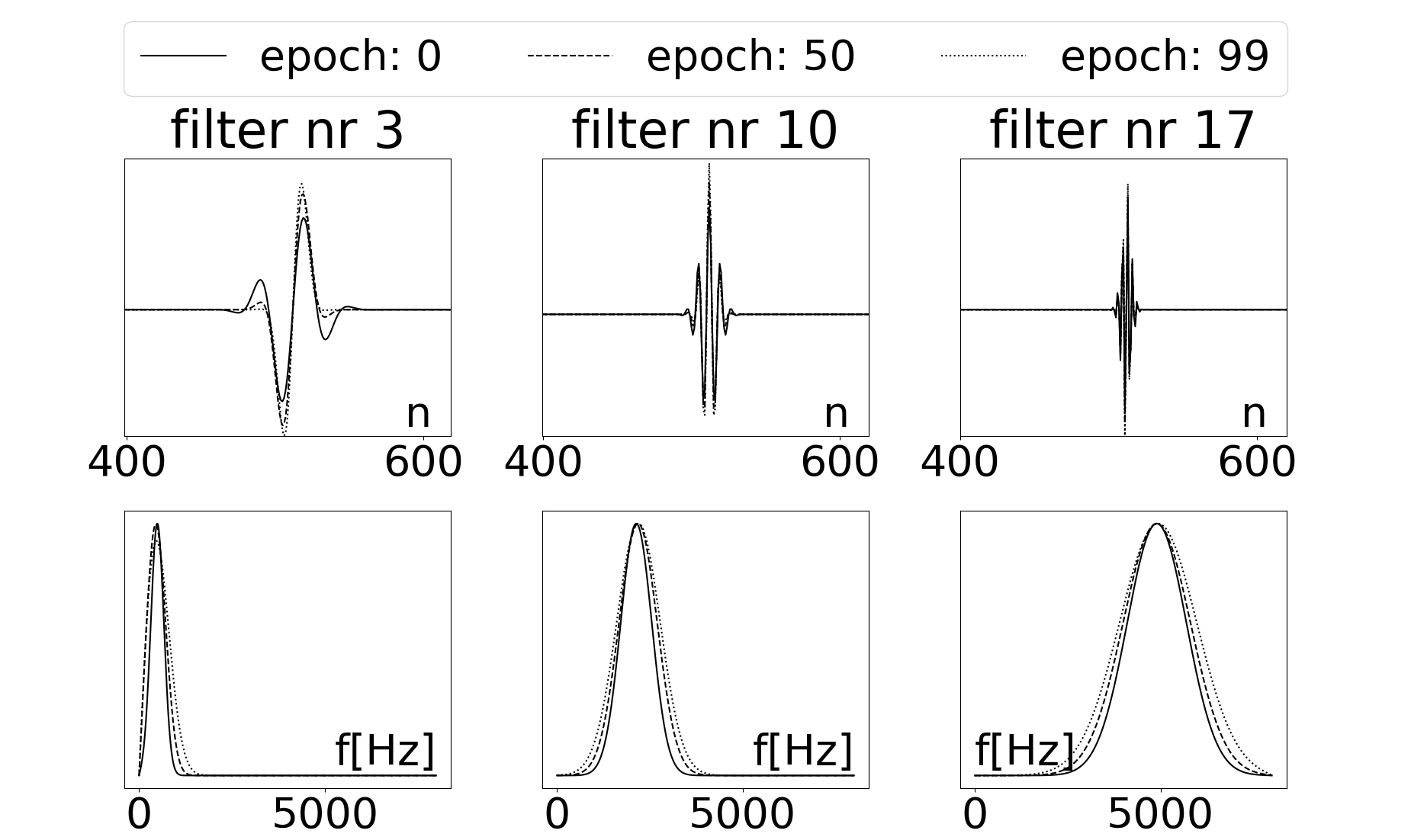}
        \caption{Three examples of Gabor filters at the training stage across three epochs:  $0^{th}$, $50^{th}$ and $100^{th}$. The top row shows characteristics in the time domain, and the bottom row presents characteristics in the frequency domain.}
        \label{fig:fig2}
\end{figure}

The primary changes focus on the input layer, where the SincNet front end was replaced with a bank of Gabor filters. Because this substitution produces complex-valued outputs, additional adjustments were required. The spectral characteristics of the applied filters in the ingestion layer are compared in Figures \ref{fig:fig1} and \ref{fig:fig2}. Since the proposed solution is inspired by the LEAF architecture, the description begins with the algorithms used in that framework.

\subsection{LEAF}
\textit{Filterbank}. The ingestion layer convolves the input signal with a bank of complex Gabor filters. The Gabor filter bank is initialized to approximate a mel-scaled triangular filter bank \cite{8462015}. The coefficients of the \textit{t}-th filter are given by:

$$ w_t = e^{-i2\pi\eta_nt} \frac{1}{\sqrt[]{2\pi} \sigma_n} e^{-\frac{t^2}{2\sigma^2_n}} $$

$$ {where\ t} = -\frac{W}{2}, \ldots , \frac{W}{2} , n = 1,\ldots ,N $$

\noindent and $\eta_n$ is the center frequency, $\sigma_n$ is the inverse bandwidth, and \textit{W} is each filter length. A filter bank of \textit{N} filters is parameterized by a trainable vector with \textit{N} center frequencies.
\bigskip

\noindent\textit{Squared modulus}. To avoid manipulating complex-valued vectors, which are the outcome of the convolution, the N-convolved signal is squared element-wise, obtaining a real-valued sequence. This squared-modulus output can be interpreted as a set of subband Hilbert envelopes, which are invariant to small time shifts \cite{Zeghidour2021LEAFAL}.

\noindent\textit{Gaussian Lowpass Pooling}. The subsequent step is Gaussian low-pass pooling with a stride of 3, which downsamples the filterbank output. This operation is a type of low-pass filtration applied to each channel independently, preserving the distinct characteristics learned by each channel. Thus, low-pass filters of length P are parameterized to have a Gaussian impulse given by \cite{Zeghidour2021LEAFAL}:

$$ p_t = \frac{1}{\sqrt[]{2\pi}}e^{-\frac{t^2}{2\sigma^2_p}} \: p = \frac{-P}{2}, \ldots , \frac{P}{2} $$

\noindent where $\sigma_p$ is an inverse bandwidth

\bigskip

\noindent\textit{Learning Per-Channel Compression and Normalization (PCEN)}. It is type of learnable normalization. The output sequence \textit{yt} is described by:

$$ y_t = \left(\frac{F(X_{t,n})}{(\varepsilon + m_{t,n})^{\alpha_n}}  + \delta_n \right)^{r_n} - \delta^{r_n}_n$$

\noindent where \textit{t} denotes the time index and \textit{n} the channel index. The parameter $\alpha_n$ is the channel-dependent smoothing coefficient, $\varepsilon$ is a small constant added to prevent division by zero, and $\delta_n$ is a channel-dependent offset. The exponent $r_n$ in the range [0, 1] controls the amount of compression. The time-frequency representation \textit{F} of the signal is normalized by the exponential moving average of past values $m_{t,n}$:

$$ m_{0,n} = x_{0,n}, \; m_{t,n}=(1-s)m_{t-1,n} + sx_{t,n}$$

\noindent what is equal to the impulse response of an IIR filter. The variable \textit{s} is the smoothing coefficient. In the LEAF solution, all parameters are learned jointly \cite{Zeghidour2021LEAFAL}\cite{Wang2016TrainableFF}.

\subsection{Gabor RawNet2}\label{subsec2}

The first baseline model used in the current experiments which serves as the basis for further modifications, was RawNet2 \cite{9414234}\cite{Jung2020ImprovedRW}. It is a type of deep neural network that allows features to be extracted directly from raw waveforms. 

\begin{table}[hb!]
\centering
\caption{The Gabor RawNet2 architecture. Numbers given in the Conv and GaborConv layers denote the
kernel size and filter length. BN stands for Batch Normalization. SELU stands for Scaled Exponential Linear Unit. FC stands for
Fully Connected layer. FMS stands for Filter-Map Scaling as
described in \cite{Jung2020ImprovedRW}. }\label{tab:tab1}
\small
\begin{tabular}{cccc}
\textbf{Layer}                                                             & \textbf{Input: 64600 samples} & \multicolumn{2}{c}{\textbf{Output}}              \\ \hline
\multirow{6}{*}{GaborNet}                                                  & GaborConv(1024, 20)           & \multicolumn{2}{c}{\multirow{6}{*}{(20, 21192)}} \\
                                                                           & GaussianPooling(1024)         & \multicolumn{2}{c}{}                             \\
                                                                           & SqrtModActivation             & \multicolumn{2}{c}{}                             \\
                                                                           & PCEN                          & \multicolumn{2}{c}{}                             \\
                                                                           & MaxPooling(3)                 & \multicolumn{2}{c}{}                             \\
                                                                           & BN+SeLU                       & \multicolumn{2}{c}{}                             \\ \hline
\multirow{5}{*}{\begin{tabular}[c]{@{}c@{}}Residual \\ block\end{tabular}} & BN+LeakyReLU                  & \multirow{9}{*} & \multirow{9}{*}{(128,784)} \\
                                                                           & Conv(3, 1, 20)                & { \rdelim\}{6}{*} [ $\times3$]}                       &                            \\
                                                                           & BN+LeakyReLU                  &                     &                            \\
                                                                           & Conv(3, 1, 20)                &                     &                            \\
                                                                           & MaxPool(3)                    &                     &                            \\ \cmidrule{1-2} 
\multirow{4}{*}{FMS}                                                       & AdaAvgPool(1)                 &                     &                            \\
                                                                           & FC + Sigmoid                  &                     &                            \\
                                                                           & Filter-wise rescale           &                     &                            \\ \hline
\multirow{5}{*}{\begin{tabular}[c]{@{}c@{}}Residual \\ block\end{tabular}} & BN+LeakyReLU                  & \multirow{9}{*} & \multirow{9}{*}{(128,29)}  \\
                                                                           & Conv(3, 1, 128)               & { \rdelim\}{6}{*} [ $\times3$]}                       &                            \\
                                                                           & BN+LeakyReLU                  &                     &                            \\
                                                                           & Conv(3, 1, 128)               &                     &                            \\
                                                                           & MaxPool(3)                    &                     &                            \\ \cmidrule{1-2}
\multirow{4}{*}{FMS}                                                       & AdaAvgPool(1)                 &                     &                            \\
                                                                           & FC + Sigmoid                  &                     &                            \\
                                                                           & Filter-wise rescale           &                     &                            \\ \hline
\multirow{2}{*}{GRU}                                                       & BN, SeLU                      & \multicolumn{2}{c}{\multirow{2}{*}{(1024)}}      \\
                                                                           & GRU(128, 1024)                & \multicolumn{2}{c}{}                             \\ \hline
FC1\footnotemark[4]                                                         & FC(1024)                      & \multicolumn{2}{c}{(1024)}                       \\ \hline
FC2                                                                        & FC(2)                         & \multicolumn{2}{c}{(2)}                          \\ \hline
\textbf{Output}                                                            & \textbf{LogSoftMax}           & \multicolumn{2}{c}{\textbf{(2)}}

\end{tabular}
\end{table}

Originally, the first layer was a SincNet, consisting of a one-dimensional time-convolution layer whose filter parameters follow a sinc function. The filter centers were initialized using a mel filter bank. These filters were parameterized by the learnable minimum and maximum cutoff frequencies \cite{8639585}. In the current solution, the sinc functions in the ingestion layer were replaced with Gabor filters, each having learnable center frequencies and bandwidths \cite{Zeghidour2021LEAFAL}. The convolved signal is then squared element-wise, resulting in a real-valued sequence. 
To downsample the obtained signal, Gaussian low-pass pooling was applied. Next, six residual blocks were used, which were left unchanged relative to the baseline architecture. At the output of each residual block, filter-wise feature map scaling (FMS) was applied independently.
FMS adjusts the scale of a given feature map, which is the output of a residual block, to produce a more discriminative representation.  This allows the neural network to be optimized to place greater attention to sub-bands that FMS identifies as more informative. 

\begin{figure}[h]
    \centering
    \includegraphics[width=0.3\textwidth]{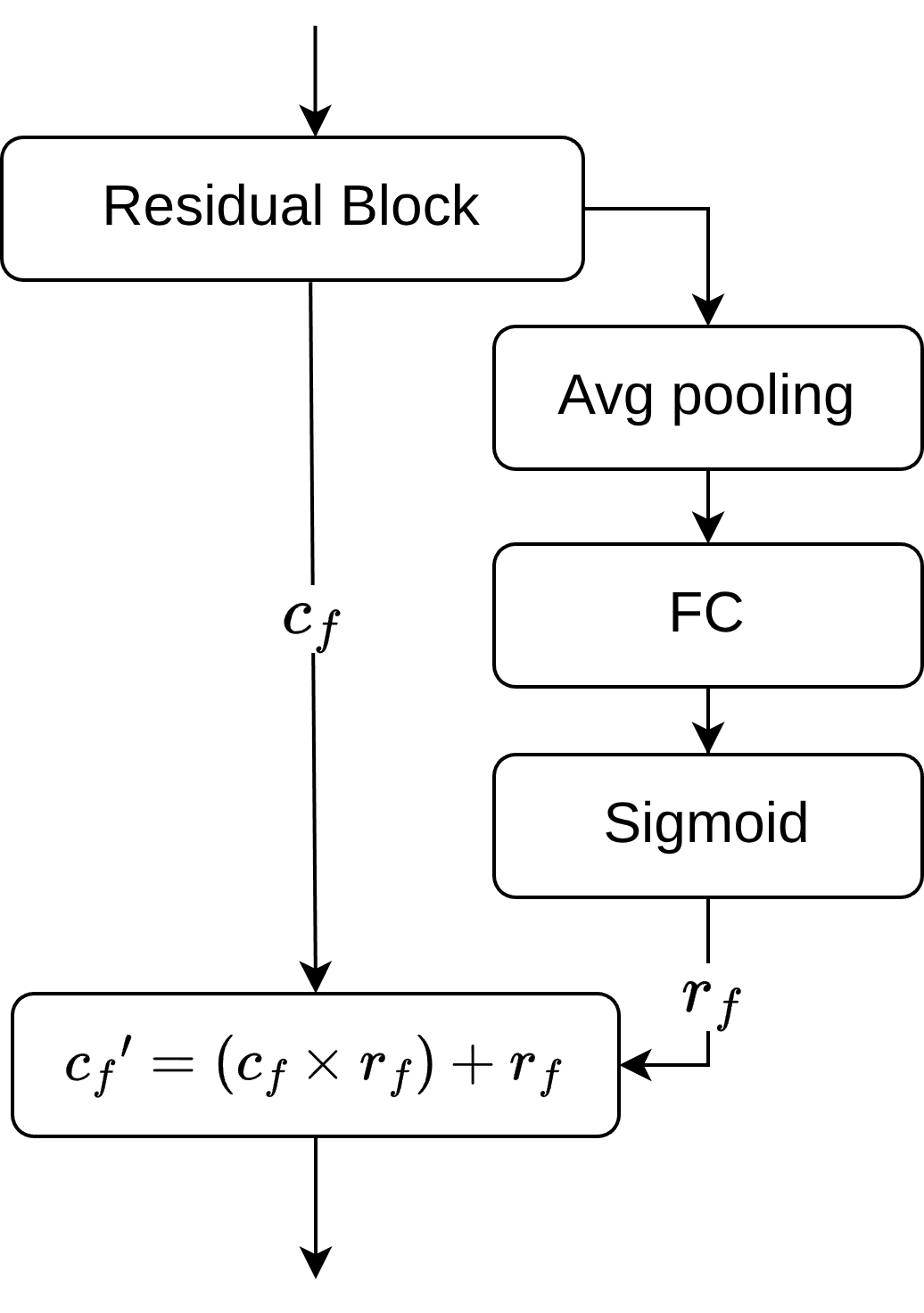}
    \caption{The schema of Filter Map Scaling applied in Gabor RawNet2.}
    \label{fig:fig3}
\end{figure}

To rescale a given feature map, various FMS techniques were investigated, including additive and multiplicative methods. However, it has been observed that the best results are achieved using $c\text{'}_{f}=(c_f\times r_f)+r_f$  (Fig \ref{fig:fig3}), where $c_f$ is an output of a residual block and $r_f$ is a rescale map derived from $c_f$ by performing average pooling along the time axis, followed by a fully connected network with a sigmoid activation function \cite{Jung2020ImprovedRW}.

The final stage remains unchanged compared to the baseline architecture and consists of the following elements: a Gated Recurrent Unit (GRU) with 1024 nodes, two Fully Connected (FC) layers, and a softmax activation function that produces a two-class results: spoof or bona-fide. The proposed modified version of the RawNet2 architecture, which we'll refer to as Gabor RawNet2, is presented in Table \ref{tab:tab1}.

\subsection{Gabor RawGAT-ST}\label{subsec3}

Another architecture used here as the host of the ingestion layer, consisting of a Gabor filter bank, is the RawNet Spectro-Temporal Graph Attention Network (RawGAT-ST). In the original solution, the role of the ingestion layer was played by SincNet. In the current study, it was replaced by Gabor filters with learnable center frequencies and bandwidths for each filter in the bank \cite{Zeghidour2021LEAFAL}. Next, there are six residual blocks, which were left unchanged compared to the baseline architecture of Hammleta et al. \cite{Tak2021EndtoEndSG}. 

The output of the residual blocks is a high-level feature map with three dimensions (channels × frequency × time). The temporal and spectral blocks separately reduce dimensions by max pooling. In the temporal branch, pooling is applied along the spectral dimension, yielding a temporal feature map (channels × time). In the spectral branch, pooling is applied along the temporal dimension, yielding a spectral feature map (channels × frequency).

\begin{table}[h]
\centering
\caption{Gabor RawGAT-ST architecture.}\label{tab:tab2}
\small
\begin{tabular}{cccc}
\textbf{Layer}                                                                         & \textbf{Input: 64600 samples}                                                                          & \multicolumn{2}{c}{\textbf{Output}}                                                                             \\ \hline
\multirow{6}{*}{GaborNet}                                                              & GaborConv(128, 70)                                                                                     & \multirow{6}{*}{}                                 & \multirow{6}{*}{(23, 21490)}                                \\
                                                                                       & GaussianPooling(128)                                                                                   &                                                   &                                                             \\
                                                                                       & SqrtModActivation                                                                                      &                                                   &                                                             \\
                                                                                       & PCEN                                                                                                   &                                                   &                                                             \\
                                                                                       & MaxPooling(3)                                                                                          &                                                   &                                                             \\
                                                                                       & BN+SeLU                                                                                                &                                                   &                                                             \\ \hline
\multirow{8}{*}{\begin{tabular}[c]{@{}c@{}}Residual \\ block\end{tabular}}             & BN                                                                                                     & \multirow{8}{*}                               & \multirow{8}{*}{(32, 23, 795)}                              \\
                                                                                       & Conv2d((2, 3), 32)                                                                                     &  { \rdelim\}{5}{*} [ $\times2$]}                                                  &                                                             \\
                                                                                       & SeLU                                                                                                   &                                                   &                                                             \\
                                                                                       & Conv2d((2, 3), 32)                                                                                     &                                                    &                                                             \\
                                                                                       & BN + SeLU                                                                                                     &                                                   &                                                             \\
                                                                                       & Conv2d((2, 3), 32)                                                                                     &                                                   &                                                             \\
                                                                                       & MaxPool(1, 3)                                                                                          &                                                   &                                                             \\ \hline
\multirow{8}{*}{\begin{tabular}[c]{@{}c@{}}Residual \\ block\end{tabular}}             & BN                                                                                                     & \multirow{8}{*}                             & \multirow{8}{*}{(64, 23, 29)}                               \\
                                                                                       & Conv2d((2, 3), 64)                                                                                     &  { \rdelim\}{5}{*} [ $\times4$]}                                                  &                                                             \\
                                                                                       & SeLU                                                                                                   &                                                   &                                                             \\
                                                                                       & Conv2d((2, 3), 64)                                                                                     &                                                   &                                                             \\
                                                                                       & BN + SeLU                                                                                                      &                                                   &                                                             \\
                                                                                       & Conv2d((2, 3), 32)                                                                                     &                                                   &                                                             \\
                                                                                       & MaxPool(1, 3)                                                                                          &                                                   &                                                             \\ \hline
                                                                                       & Spectral Branch                                                                                        & \multicolumn{2}{c}{Temporal Branch}                                                                             \\ \hline
Attention                                                                              & \begin{tabular}[c]{@{}c@{}}$max_{T}$(64, 23)\\ GAT layer (32, 23)\end{tabular}                         & \multicolumn{2}{c}{\begin{tabular}[c]{@{}c@{}}$max_{F}$(64, 29)\\ GAT layer (32, 29)\end{tabular}}              \\ \hline
\begin{tabular}[c]{@{}c@{}}Graph \\ Pooling\end{tabular}                               & \begin{tabular}[c]{@{}c@{}}FC(32) + Sigmoid\\ top-k(int{[}0.64$\times$23{]})\end{tabular}            & \multicolumn{2}{c}{\begin{tabular}[c]{@{}c@{}}FC(32) + Sigmoid\\ top-k(int{[}0.81$\times$29{]})\end{tabular}} \\ \hline
Projection                                                                             & FC(32, 12)                                                                                             & \multicolumn{2}{c}{FC(32, 12)}                                                                                  \\ \hline
\multicolumn{4}{c}{Fusion}                                                                                                                                                                                                                                                                                        \\ \hline
Multi-level                                                                            & \begin{tabular}[c]{@{}c@{}}Element-wise \\ multipication\end{tabular}                                  &                                                   & (32, 12)                                                    \\ \hline
\multirow{3}{*}{\begin{tabular}[c]{@{}c@{}}Spectro-Temporal \\ attention\end{tabular}} & GAT layer                                                                                              &                                                   & (16, 12)                                                    \\
                                                                                       & Graph-Pooling                                                                                          &                                                   &                                                             \\
                                                                                       & \begin{tabular}[c]{@{}c@{}}FC(16)\\ Sigmoid\\ top-k (int{[}0.64$\times$23{]})\\ Drop(0.3)\end{tabular} &                                                   & (16, 7)                                                     \\ \hline
Projection                                                                             & FC(16)                                                                                                 &                                                   & (1, 7)                                                      \\ \hline
\textbf{Output}                                                                        & \textbf{FC(2)}                                                                                         & \textbf{}                                         & (2)                                                        
\end{tabular}
\end{table}

\begin{figure}[h]
    \centering
    \includegraphics[width=0.3\textwidth]{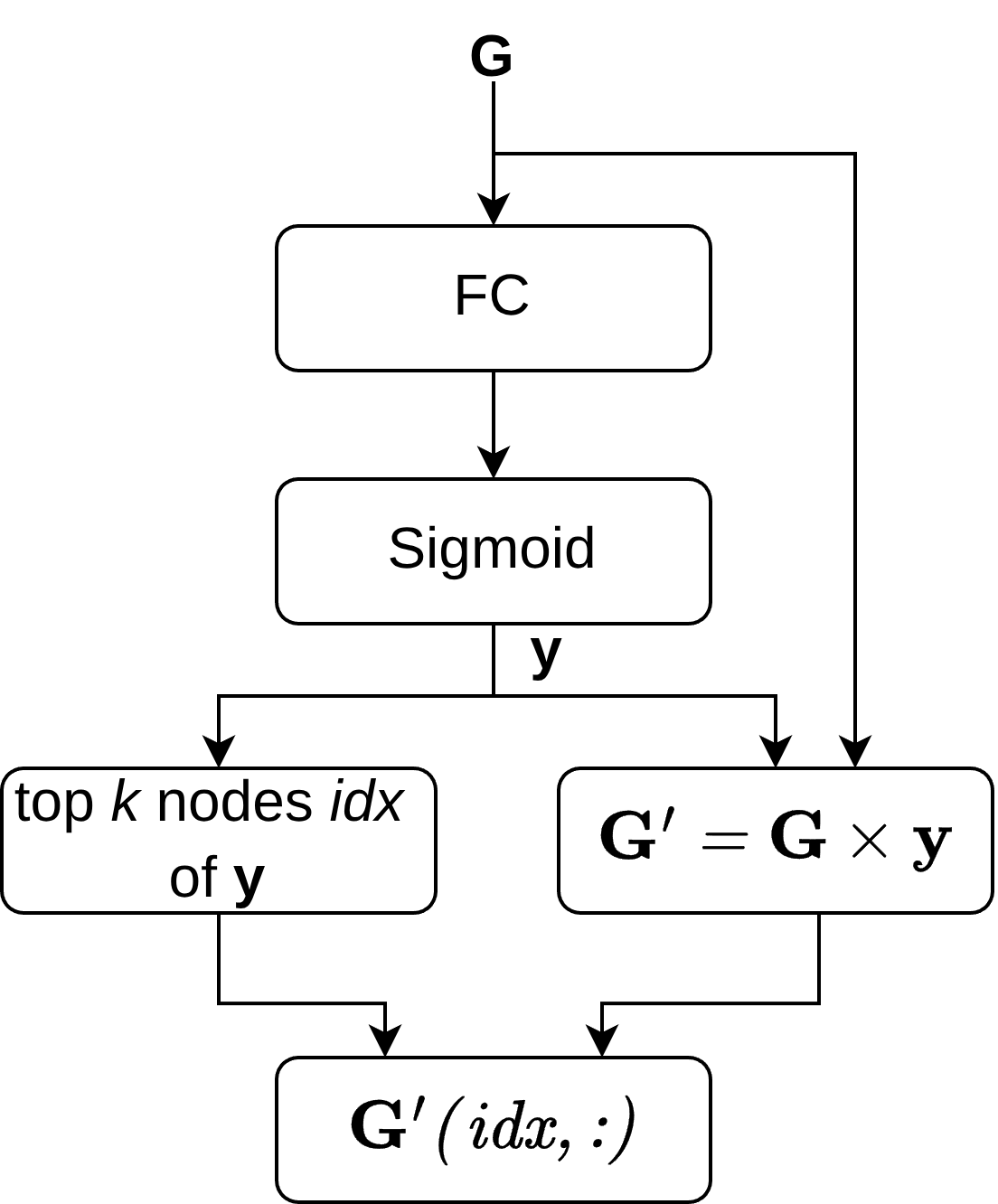}
    \caption{The schema of the Top K-Pooling layer applied in RawGAT-ST separately to time, frequency and fusion domains.}
    \label{fig:fig4}
\end{figure}

The outputs of these branches serve as inputs to two separate Graph Attention Layers. The Spectral-GAT ($G_f$) processes a set of 23 spectral bins, whereas the Temporal-GAT ($G_t$) processes a set of 29 temporal blocks. These two parallel, separate GAT-ST layers share a common channel dimension of 32.
After each GAT block, a pooling layer is applied. Pooling layers can reduce the size of feature maps, remove redundant information, and emphasize regions with strong discriminative power, leading to better generalization and improved performance. In grid-like data, pooling methods are typically based on dividing feature maps into uniform regions and taking a single value from each region using one of the following criteria: maximum, mean, or stochastic selection. However, such pooling operations cannot be directly applied to graph-structured data \cite{9432709}. In this study, the Top K-Pooling method (Fig. \ref{fig:fig4}), a solution dedicated to Graph Neural Networks, was used \cite{Tak2021GraphAN}\cite{9432709}\cite{Knyazev2019UnderstandingAA}. 

First, the sigmoid function is applied, producing the tensor \textit{y}. Next, the dot product of each input structure \textit{G} and \textit{y} yields projection matrices \textit{G'}. In the following step, the indices of the \textit{k} largest elements of the tensor \textit{y} along the node dimension are determined. For example, if the predefined hyperparameter \textit{k} equals 0.8, only 80{\%} of the nodes are retained after each pooling layer. The returned indexes (idx) contains the indices of nodes selected for the new graph. Node extraction is performed by slicing {\textit{G'}}(idx, :).
Finally, fusion of the spectral and temporal graphs is performed by multiplication: $G_{f,pooled}' \times  G_{t,pooled}'$. In the last stage, a Graph Attention Layer and another Top K-Pooling layer are applied. The final binary prediction (bona fide or spoofed) is obtained using a fully connected layer with a two-dimensional output. The proposed modified RawGAT-ST architecture schema is presented in table \ref{tab:tab2} and is named Gabor RawGAT-ST. 

\subsection{Augmentation methods}

In this study, several methods of audio augmentation in the context of audio spoof detection were investigated. The first method uses the Room Impulse Response Dataset (RIR) and the Music, Speech, and Noise Corpus (MUSAN) \cite{Cohen2021ASO}\cite{Snyder2015MUSANAM}. In the case of room responses from the RIR dataset, these were introduced to the audio through convolution. On the other hand, to introduce distortions from MUSAN to the training audio, the operation of addition was applied. The second exploited method was based on distortions from common audio compression algorithms such as aLaw, uLaw, MP3, G.727, and Ogg. The applied lossy compression algorithms were derived from Torchaudio Sox and Ffmpeg backends \footnote{https://docs.pytorch.org/audio/main/torchaudio.html}. Experiments were also conducted with SpecAugment methods, but these were found to be ineffective in the spoof detection task; therefore, results were not presented here \cite{Park2019SpecAugmentAS}.

\begin{table}[h!]
\centering
\caption{Procedure for Training with Audio Augmentation at Each Epoch Using RIR and MUSAN Datasets \cite{Cohen2021ASO}\cite{Snyder2015MUSANAM}}\label{tab:tab3}
\begin{tabular}{rllll}
signal      & \multicolumn{3}{l}{$\leftarrow$ array{[}0:64600{]}}                    &               \\
num         & \multicolumn{3}{l}{$\leftarrow$ random(0, 5)}                          &               \\ \hline
\textbf{if} & num=0   & \textbf{then}               &                                &               \\
            &         & \textbf{return}             & signal                         &               \\ \hline
\textbf{if} & num=1   & \textbf{then}               &                                &               \\
            &         & \multicolumn{2}{l}{rir $\leftarrow$ choice(RIRs files list)} &               \\
            &         & \textbf{return}             & \textbf{convolution}  (signal, rir)         &  \\ \hline
\textbf{if} & num=2   & \textbf{then}               &                                &               \\
            &         & \multicolumn{3}{l}{musan $\leftarrow$ choice(MUSAN-speech)}                  \\
            &         & \textbf{return}             & signal+musan                   &               \\ \hline
\textbf{if} & num=3   & \textbf{then}               &                                &               \\
            &         & \multicolumn{3}{l}{musan $\leftarrow$ choice(MUSAN-music)}                   \\
            &         & \textbf{return}             & signal+musan                   &               \\ \hline
\textbf{if} & num=4   & \textbf{then}               &                                &               \\
            &         & \multicolumn{3}{l}{musan $\leftarrow$ choice(MUSAN-noise)}                   \\
            &         & \textbf{return}             & signal+musan                   &               \\ \hline
\textbf{if} & num=5   & \textbf{then}               &                                &               \\
            &         & \multicolumn{3}{l}{musan $\leftarrow$ choice(MUSAN-speech)}                  \\
            &         & \multicolumn{3}{l}{musan $\leftarrow$ musan+choice(MUSAN-music)}             \\
            &         & \textbf{return}             & signal+musan                   &              
\end{tabular}
\end{table}

\section{Experimental setup}\

In this section, the databases and evaluation metrics used for the experiments are described. Additionally, further details of the investigated architectures are presented.

\subsection{Database and evaluation metrics}

Experiments were performed on the ASVspoof 2019 Logical Access database. This is part of the ASVspoof 2019 dataset based on the Voice Cloning Toolkit (VCTK) corpus of English speakers. The LA (Logical Access) scenario imitates an attempt at remote access to resources protected by an automatic speaker verification system. "Remotely" means that the user's microphone is outside the authentication system's control. The LA dataset has three independent subsets: training, validation, and evaluation. The training set contains speech recordings of 20 speakers (8 male and 12 female), while the validation set contains 10 speakers (4 male, 6 female) \cite{WANG2020101114}. Training and validation non-target samples were generated using 6 different TTS or VC systems. The evaluation set consists of speech samples from 48 bonafide speakers (21 male, 27 female) and 19 spoofed speakers (10 female, 9 male). To generate the synthetic or converted speech of the evaluation set, 12 TTS or VC systems were used, marked as A07 - A19, which were not seen during the training stage \cite{WANG2020101114}.
To assess the results of experiments, the Equal Error Rate (EER) was used. The well-known EER is a point on the error rate function of the threshold (operating point), such as ROC (Receiver Operating Characteristic) or DET (Detection Error Tradeoff), where the probability of false acceptance and missed detection are approximately equal \cite{Brmmer2013TheBT}.

\subsection{Baseline}\label{subsec6}

\begin{figure}[h]
    \centering
    \includegraphics[width=0.5\textwidth]{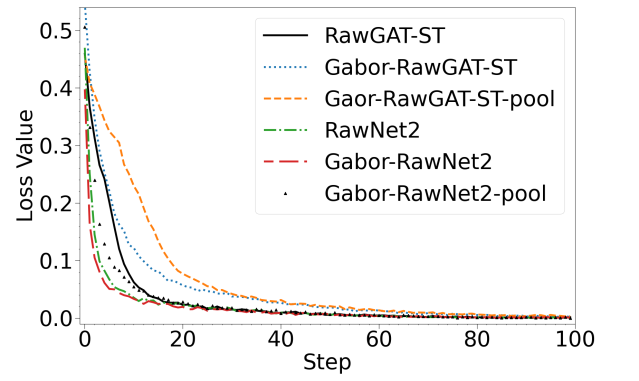}
    \caption{Comparison of training curves of investigated architectures.}
    \label{fig:fig5}
\end{figure}

Experiments have been started with state-of-the-art architectures, RawNet2 and RawGAT-ST. This initial phase allows us to ensure consistent conditions throughout the investigation by determining the baseline results. In the case of both examined architectures, filter banks with sinc functions were initialized with 20 and 32 mel-scaled filters, respectively. SincNet was utilized as the ingestion layer with kernel sizes of 1024 and 128, respectively, and stride equal to 1. In the second stage of experiments, RawNet2 and RawGAT-ST were adapted by replacing SincNet with GaborNet as the ingestion layer. The complex Gabor filter bank was initialized with 70 mel-scaled filters. Kernel sizes equal 1024 and 128, respectively, and stride equals 1. The Adam optimizer was used with a base learning rate of 0.0001 and an adaptive adjustment option during optimization via a cosine learning rate scheduler. A comparison of the training curves of various modifications of the investigated architectures is presented in Figure \ref{fig:fig5}.

\subsection{Results}\label{subsec8}

The obtained results are illustrated in Tables \ref{tab:tab4}, \ref{tab:tab5}, and \ref{tab:tab6}, where all columns indicate EER. Table \ref{tab:tab5} consists of three groups of outcomes:
\begin{enumerate}
    \item results of baseline architectures (RawNet2 and RawGAT-ST)
    \item results of using the GaborNet ingestion 	layer in place of SincNet (named 	Gabor-RawNet2 and Gabor-RawGAT-ST)
    \item results of the complete LEAF solution used as an ingestion layer with Gabor Convolution layer, Gaussian Lowpass Pooling, and PCEN (named LEAF-RawNet2 and LEAF-RawGAT-ST)   
\end{enumerate}

Table \ref{tab:tab6} presents the results of baseline architectures trained with data augmentation using three methods, abbreviated as follows: codec – audio compression distortions, RIR – signal distorted by acoustic room response and additive noise (MUSAN), and RIR+codec – a combination of the two previous methods.
Table \ref{tab:tab7} presents the results of LEAF-RawNet2 and LEAF-RawGAT-ST trained with data augmented by the three aforementioned methods.
Modifications to RawNet2 that include replacing SincNet with GaborNet lead to improvements in terms of EER. The original version yields 4.131{\%}, while the modified version achieves 4.025{\%}. Attaching Gaussian Low-Pass Pooling brings a further decrease in error to 3.807{\%}. However, such improvement has not been observed in the case of RawGAT-ST. Applying GaborNet has yielded a regression from 1.778{\%} in the original architecture to 2.000{\%}. Attaching Gaussian Low-Pass Pooling has caused a further increase in the Equal Error Rate to 2.406{\%}.
The application of data augmentation through codec conversions has achieved improvements for RawNet2, achieving an EER of 3.073{\%}. This augmentation method performs best for RawGAT-ST as well; however, the error has increased compared to the baseline training without augmentation. Other augmentation methods such as RIR+MUSAN and CODEC+RIR+MUSAN have proven to be ineffective in the investigated cases.
\begin{table}[h]
\centering
\caption{Ablation study LEAF-RawGAT-ST}\label{tab:tab7}
\begin{tabular}{c|c}
Type of nn                                  & EER    \\ \hline
LEAF (GaborNet +GaussianPooling+PCEN)       & 21.588 \\
SincNet                                     & 50.116 \\
GaborNet PCEN w/o GaussianPooling           & 19.485 \\
LEAF-RawGAT-S (w/o Temporal GAT and Fusion) & 6.787  \\
LEAF-RawGAT-T (w/o Spectral GAT and Fusion) & 1.996  \\
GaborNet-RawGAT-T (w/o Fusion)              & 2.788  \\
GaborNet-RawGAT-ST                          & 1.778 
\end{tabular}
\end{table}

\subsection{Ablation study}\label{subsec7}

Results show that using Gabor convolution and a more complex LEAF architecture in the role of the ingestion layer, in the case of RawNet2, contributes to reducing EER. For the RawGAT-ST architecture, the opposite effect is seen. To explain which part of the deep neural network has the greatest impact on performance, an ablation study has been conducted. Using only the Learnable Frontend for Audio Classification (LEAF) leads to an error of 21.588{\%}, meanwhile using standalone SincNet leads to 50.116{\%} EER. This means that the LEAF neural network extracts much more valuable information for audio spoof detection than the commonly used SincNet. Results of abblation study are presented in Table \ref{tab:tab7}.

\subsection{Conclusion}

Conducted experiments show that Learnable Frontend for Audio Classification (LEAF) can be used as an effective ingestion layer in audio spoof detection neural network architectures based on RawNet2 solutions. The obtained results also show that LEAF in the role of the first layer is not efficient with RawGAT-ST. Conducted ablation experiments show that the LEAF front-end and Spectral Graph Attention mechanism have the most significant contribution to performance degradation.
The best augmentation method appears to be transcoding using various lossy compressions, such as A-law, $\mu$-law, MP3, G.727, and Ogg. Other methods, like convolving the signal with acoustic room impulse responses (RIR) or adding noise from the MUSAN dataset (such as speech and music), as well as combinations of the above methods under the conditions considered, have appeared to be ineffective.

\bigskip

\begin{table*}[h]
\centering
\caption{Results in terms of EER of RawNet2, Gabor-RawNet2, RawGAT-ST, Gabor-RawGAT-ST and LEAF-RawGAT-ST on the ASV Spoof 2019 LA dataset.}\label{tab:tab4}
\begin{adjustbox}{width=0.99\textwidth}
\small
\begin{tabular}{c||cccccccccccccc}
Type of nn      & A07            & A08            & A09            & A10            & A11           & A12            & A13            & A14            & A15            & A16            & A17            & A18            & A19            & all            \\ \hline
RawNet2         & 3.952          & 5.208          & 0.221          & 4.254          & 3.131         & 4.801          & 0.547          & 1.158          & 3.253          & 1.199          & 6.244          & 9.242          & 1.630          & 4.131          \\
Gabor-RawNet2   & 4.540          & 4.074          & 0.587          & 4.271          & 1.786         & 4.271          & 0.424          & \textbf{0.228} & 3.905          & \textbf{0.913} & 3.579          & 17.52          & 1.117          & 4.025          \\
LEAF-RawNet2    & 4.604          & 4.662          & 0.8387         & 6.285          & 4.981         & 6.146          & 0.954          & 1.280          & 4.604          & 1.63           & 2.299          & 10.01          & 1.565          & 3.807          \\
RawGAT-ST       & 3.300          & 1.200          & 0.506          & 4.312          & 1.874         & 4.523          & 2.462          & 1.426          & 4.54           & 1.443          & 1.728          & \textbf{4.149} & 1.321          & \textbf{1.778} \\
Gabor-RawGAT-ST & 3.195          & 0.954          & \textbf{0.099} & 4.638          & 2.299         & 4.703          & 1.484          & 0.465          & 2.730          & 1.402          & \textbf{1.589} & 4.394          & \textbf{0.815} & 2.000          \\
LEAF-RawGAT-ST  & \textbf{2.037} & \textbf{0.913} & 0.163          & \textbf{3.986} & \textbf{1.63} & \textbf{3.864} & \textbf{0.309} & 0.448          & \textbf{1.915} & 1.141          & 2.479          & 6.798          & 1.525          & 2.406         
\end{tabular}
\end{adjustbox}
\end{table*}
\begin{table*}[h]
\centering
\caption{ Results in terms of EER for RawNet2 and RawGAT-ST on the augmented training set from ASVspoof 2019 LA, using lossy compression (Codec), room impulse response, and/or additive distortions (RIR).}\label{tab:tab5}
\begin{adjustbox}{width=0.99\textwidth}
\small
\begin{tabular}{c||cccccccccccccc}
Type of nn            & A07            & A08            & A09            & A10            & A11            & A12            & A13            & A14            & A15            & A16            & A17            & A18            & A19            & all            \\ \hline
RawNet2 Codec         & \textbf{0.407} & \textbf{0.343} & 5.429          & 0.221          & \textbf{0.513} & 0.553          & 0.424          & 0.367          & \textbf{0.390} & 0.513          & 5.103          & 8.835          & 1.239          & 3.073          \\
RawNet2 RIR           & 0.978          & 0.587          & 3.521          & 0.187          & 0.856          & 0.465          & 0.424          & \textbf{0.367} & 0.676          & \textbf{0.995} & 11.337         & 22.878         & 3.708          & 6.485          \\
RawNet2 RIR+Codec     & 0.733          & \textbf{0.343} & 4.964          & 0.204          & 0.587          & 0.407          & 0.244          & 0.407          & 0.431          & \textbf{0.733} & 9.847          & 20.549         & 2.910          & 6.077          \\
RawGAT-ST Codec       & 2.788          & 0.448          & 0.937          & 0.350          & 0.628          & \textbf{0.204} & \textbf{0.180} & 0.448          & 0.553          & 1.385          & \textbf{3.416} & \textbf{5.610} & \textbf{1.222} & \textbf{2.094} \\
RawGAT-ST RIR         & 0.856          & 0.669          & \textbf{0.815} & \textbf{0.139} & 0.954          & 0.465          & 0.302          & \textbf{0.139} & 0.594          & 0.995          & \textbf{9.137} & 16.749         & \textbf{1.246} & 4.337          \\
RawGAT-ST RIR + Codec & \textbf{2.723} & \textbf{1.833} & 0.856          & \textbf{0.146} & \textbf{3.619} & \textbf{1.222} & \textbf{0.937} & 0.221          & \textbf{1.182} & 2.159          & 10.481         & 8.852          & 2.747          & 4.062         
\end{tabular}
\end{adjustbox}
\end{table*}

\begin{table*}[h]
\centering
\centering
\caption{Results in terms of EER for LEAF-RawNet2 and LEAF-RawGAT-ST using an augmented training set from ASVspoof 2019 LA with lossy compression (Codec), room impulse response, and/or additive distortions (RIR).}\label{tab:tab6}
\begin{adjustbox}{width=0.99\textwidth}
\small
\begin{tabular}{c|cccccccccccccc}
Type of nn + augmentation & A07            & A08            & A09            & A10            & A11            & A12            & A13            & A14            & A15            & A16            & A17            & A18            & A19            & all            \\ \hline
LEAF-Raw Net2 Codec       & 3.056          & 2.869          & 4.475          & 1.117          & 2.886          & 8.200          & 4.278          & 2.805          & 1.752          & 2.013          & 12.478         & 28.755         & 2.706          & 7.750          \\
LEAF-RawNet2 RIR          & 2.078          & 1.728          & 3.585          & 0.710          & 1.745          & 1.117          & 1.263          & 1.402          & 1.222          & 1.769          & 18.315         & 31.241         & 1.630          & 7.928          \\
LEAF-Raw Net2RIRCodec     & 2.706          & 5.103          & 1.100          & 3.171          & 3.806          & 4.394          & 2.682          & 5.674          & 2.788          & 2.852          & 24.745         & 28.976         & 2.828          & 9.561          \\
LEAF-RawGAT-ST Codec      & \textbf{0.326} & \textbf{0.390} & 0.390          & 0.350          & \textbf{0.489} & 0.530          & 0.390          & 0.343          & \textbf{0.343} & 0.710          & \textbf{4.842} & \textbf{6.227} & 0.733          & \textbf{2.406} \\
LEAF-RawGAT-ST RIR        & 2.445          & 0.407          & \textbf{0.367} & 0.343          & 1.630          & 0.587          & 0.587          & 0.122          & 0.978          & \textbf{0.611} & 6.251          & 11.093         & \textbf{0.390} & 3.482          \\
LEAF-RawGAT-ST RIR+Codec  & 0.791          & 0.547          & 0.570          & \textbf{0.105} & 0.716          & \textbf{0.261} & \textbf{0.146} & \textbf{0.122} & 0.448          & 1.246          & 6.944          & 8.509          & 0.815          & 3.100         
\end{tabular}
\end{adjustbox}
\end{table*}

\bigskip

\printcredits

\bibliographystyle{cas-model2-names}

\bibliography{cas-refs}




\vskip3pc


\end{document}